# Are Strong Brønsted Acids Necessarily Strong Lewis Acids?

K. Gupta[1,2], D. R. Roy[1], V. Subramanian[3,*] and P. K. Chattaraj[1,*]

[1]Chemistry Department, Indian Institute of Technology, Kharagpur-721302, India
[2]On leave from Department of Chemisty, Ramananda College, Bishnupur-722122, Bankura, India.
[3]Chemical Laboratory, Central Leather Research Institute, Adyar, Chennai- 600 020, India.
E-mail: pkc@chem.iitkgp.ernet.in, subuchem@hotmail.com

## Abstract

The Brønsted and Lowry acid-base theory is based on the capacity of proton donation or acceptance (in the presence/absence of a solvent) whereas the Lewis acid-base theory is based on the propensity of electron pair acceptance or donation. We explore through DFT calculation the obvious question whether these two theories are in conformity with each other. We use p$K_a$ as the descriptor for the Brønsted and Lowry acidity. The DFT descriptors like ionization potential, electron affinity, electronegativity, hardness and global electrophilicity are computed for 58 organic and inorganic acids. The fractional electron transfer, $\Delta N$ and the associated energy change, $\Delta E$ for the reaction of these acids with trimethyl amine (a strong base) are used as the possible descriptors for the Lewis acidity. A near exponential decrease in $\Delta N$ and $(-\Delta E)$ values is observed in general with an increase in p$K_a$ values. The findings reveal that a stronger Brønsted acid in most cases behaves as a stronger Lewis acid as well. However it is not necessarily true for all acids.

## Introduction

Brønsted and Lowry[1] suggested that any substance that acts as a proton donor should be classified as an acid and any substance that accepts a proton should be classified as a base. Substances that act in this way are called Brønsted acids and bases, respectively. The definitions make no reference to the environment in which proton transfer occurs, so they apply to proton transfer behavior in any solvent, and even in no solvent at all. However, by far the most important medium has been considered to be an aqueous solution and the attention is confined to that only. On the other hand, a Lewis acid[1] is a substance that acts as an electron pair acceptor. A Lewis base is a substance that acts as an electron pair donor. According to Brønsted and Lowry, a stronger acid has a smaller p$K_a$ value whereas a stronger Lewis acid has a stronger capability to accept a pair of electrons. Therefore, the ionization potential, electron affinity, electronegativity and electrophilicity may be considered to be measures of the strength of a Lewis acid. An acid with a lower p$K_a$ value is expected to have a higher electron affinity, electronegativity and electrophilicity compared to other acids of similar type if the two theories need to correlate.

Density functional theory[2-5] based descriptors may be useful in the prediction of Lewis acidity and basicity of molecules. Ionization potential (I), electron affinity (A), electronegativity ($\chi$), chemical hardness ($\eta$) and chemical potential ($\mu$) are termed as global reactivity descriptors. Parr et al.[6] have defined a new descriptor to quantify the global electrophilic power of the molecule as electrophilicity index ($\omega$), which provides a quantitative classification of the global electrophilic nature of a molecule within a relative scale. None of these parameters can singly describe Lewis acidity or basicity. Therefore, a different parameter is necessary to describe that.

The interaction process between an acid and a base is dissected into two steps : a charge transfer process resulting in a common chemical potential describing the strengths of the acid and the base, at a fixed external potential followed by a reshuffling process at a fixed chemical potential.[7] The fractional number of electron transfer, $\Delta N$ and the associated energy change, $\Delta E$ in the charge transfer depend on the interplay between electronegativity and hardness of the acid and the base[8-10] which are dependent on previously mentioned DFT descriptors. The difference in electronegativity drives electron transfer and the sum of the hardness parameters acts as a resistance.[11] An ab-initio solvation model study is performed in the recent past[12] to predict the p$K_a$ values of a few carbenes.

In the present study we compute electronegativity ($\chi$), chemical hardness ($\eta$), chemical potential ($\mu$) and global electrophilicity ($\omega$) for a series of 58 molecules (Table 1). We also compute the fractional number of electrons transferred ($\Delta N$) and energy change ($\Delta E$) for the reaction of these acids with trimethyl amine, known to be a strong base, from these parameters. These two parameters are employed as descriptors of Lewis acidity and basicity. The correlation of $\Delta N$ and $\Delta E$ with the p$K_a$ values of acids and bases are studied here for the first time.

**Theoretical Background**

Parr and co-workers[2] interpreted that chemical potential ($\mu$) could be written as the partial derivative of the system's energy with respect to the number of electrons at a fixed external potential $v(\vec{r})$ :

$$\mu = \left(\frac{\partial E}{\partial N}\right)_{v(\vec{r})} \quad (1)$$

Iczkowski and Margrave[13] proposed to define electronegativity as

$$\chi = -\left(\frac{\partial E}{\partial N}\right)_{v(\vec{r})} \quad (2)$$

for a fixed nuclear charge.

The working formulas in DFT for calculating chemical potential (µ), electronegativity (χ) and hardness (η) are as follows:

$$\mu \approx -(I+A)/2 \; ; \; \chi \approx (I+A)/2 \; ; \; \eta \approx (I-A)/2 \quad (3)$$

The ionization potential and electron affinity can be replaced by the HOMO and LUMO energies, respectively, using Koopmans' theorem[14] within a Hartree-Fock scheme yielding

$$\chi \approx -(\varepsilon_{HOMO} + \varepsilon_{LUMO})/2 \quad (4)$$

and so on.

The ionization potential and electron affinity may be better expressed as:

$$I \approx E(N-1) - E(N) \quad (5a)$$

$$A \approx E(N) - E(N+1) \quad (5b)$$

Parr et al.[6] have introduced the global electrophilicity index (ω) as a measure of energy lowering due to maximal electron flow between a donor and an acceptor in terms of the chemical potential and the hardness as

$$\omega = \frac{\mu^2}{2\eta} \quad (6)$$

If two systems, B and C, are brought together, electrons flow from that of lower χ to that of higher χ, until the chemical potentials become equal. The (fractional) number of electrons transferred for the generalized acid-base reactions
C + :B → C:B, is given (upto first order) by

$$\Delta N = \frac{(\chi_C - \chi_B)}{2(\eta_C + \eta_B)} \quad (7)$$

The energy lowering due to this electron transfer from a species of higher chemical potential (base) to that of a lower chemical potential (acid) is given by

$$\Delta E = -\frac{(\chi_C - \chi_B)^2}{4(\eta_C + \eta_B)} \quad (8)$$

The Fukui function (FF) is defined as the derivative of the electron density $\rho(\vec{r})$ with respect to the total number of electrons $N$ in the system, at constant external potential $v(\vec{r})$ acting on an electron due to all the nuclei in the system[2]

$$f(\vec{r}) = [\delta\mu/\delta v(\vec{r})]_N = [\partial\rho(\vec{r})/\partial N]_{v(\vec{r})} \quad (9)$$

where $\mu$ is the chemical potential of the system.

The generalized concept of philicity was proposed by Chattaraj et al,[15] associated with a site k in a molecule with the aid of the corresponding condensed-to-atom variants of Fukui function $f_k^\alpha$ as,[14]

$$\omega_k^\alpha = \omega \cdot f_k^\alpha \quad (10)$$

where (α = +, - and 0) represents local philic quantities describing nucleophilic, electrophilic and radical attacks. Eq. (10) predicts that the most electrophilic site in a molecule is the one providing the maximum value of $\omega_k^+$.

The group concept of philicity is very useful in unraveling reactivity of various molecular systems.[16] The condensed philicity summed over a group of relevant atoms is defined as the "group philicity". It can be expressed as

$$\omega_g^\alpha = \sum_{k=1}^{n} \omega_k^\alpha \qquad (11)$$

where n is the number of atoms coordinated to the reactive atom, $\omega_k^\alpha$ is the local electrophilicity of the atom k, and $\omega_g^\alpha$ is the group philicity obtained by adding the local philicities of the nearby bonded atoms, where ($\alpha$= +, -, 0) represents nucleophilic, electrophilic and radical attacks respectively.

**Computational Details**

The geometries of the selected series of all the 58 molecules are optimized at the B3LYP/6-31G(d) level of theory using *Gaussian 03* package.[17] The ionization potential, electron affinity, electronegativity, hardness, chemical potential and global electrophilicity index are computed employing the Koopmans' theorem[14] as well as the $\Delta$SCF method. The fractional number of the electrons transferred ($\Delta$N) and the energy change ($\Delta$E) for the reaction of these acids with trimethyl amine are computed using Eqs. (7) and (8) respectively. E(N-1) and E(N+1) are computed by single point calculations for (N-1) and (N+1)-electronic systems with the same molecular geometry obtained for the N-electronic system. Similar study is performed both in gas phase and in aqueous phase (at $298^0$K) employing the SCF energies of (N-1), N and (N+1) electronic systems. To study the solvent effects (in water medium), molecules are optimized in the framework of a self consistent reaction field polarized continuum model (PCM), using the B3LYP/6-31G(d) method. Fukui functions are calculated with the Mulliken population analysis[18] (MPA) and Hirshfeld population analysis[19] (HPA) scheme employing the BLYP/DND method using DMOL$^3$ package.[20]

## Results and Discussion

Table 1 lists the experimental p$K_a$ values,[21-25] computed electronegativity ($\chi$) and hardness ($\eta$) of 58 acids (both inorganic and organic). The table also contains the fractional number of electrons transferred, $\Delta N$ and energy change, $\Delta E$ when these acids react with trimethyl amine. Figure 1 presents the variation of experimental p$K_a$ values with the negative of the energy change associated with the electron transfer from trimethyl amine to a host of organic and inorganic acids in gas phase. A near exponential decay is easily discernible. A larger ($-\Delta E$) value implies a stronger Lewis acid and that corresponds to a smaller p$K_a$ value implying a stronger Brønsted acid. It may, however, be noted that an arbitrary pair of acids may not necessarily obey this behaviour. On an average a stronger Lewis acid is also a stronger Brønsted acid. The regression model (exponential decay) to predict p$K_a$ values using ($-\Delta E$) is as follows:

$$\text{Pred. p}K_a = 1.08(0.76) + 23.25(2.66) \times \text{EXP}[\Delta E/0.04(0.01)] \qquad (12)$$

$$R^2 = 0.749 \; ; \; N = 58$$

A reasonably good correlation between the experimental p$K_a$ value and the calculated p$K_a$ value (Table 1) is obtained. Also, corresponding regression model for the solution phase is as follows:

$$\text{Pred. p}K_a = 1.41(0.72) + 27.84(3.98) \times \text{EXP}[\Delta E/0.04(0.01)] \qquad (13)$$

$$R^2 = 0.730 \; ; \; N = 58$$

The qualitative trend does not change in the aqueous solution (Figure 2) and/or using Koopmans' theorem (not shown here). Figures 3 and 4 respectively depict the behavior of p$K_a$ with $\Delta N$ in gas and solution phases respectively. The regression models (exponential decay) to calculate p$K_a$ as a function of $\Delta$N in both the gas and the aqueous phases are given as:

Gas Phase:   Pred. p$K_a$ = -1.26(1.73) + 38.11(6.55)×EXP[–$\Delta$N/0.04(0.01)]

$$R^2 = 0.721 \; ; N = 58 \qquad (14)$$

Solution Phase: Pred. p$K_a$ = 0.92(0.85) + 82.95(24.90)×EXP[–$\Delta$N/0.03(0.01)]

$$R^2 = 0.757 \; ; N = 58 \qquad (15)$$

There exists an approximate exponential behavior between p$K_a$ and $\Delta$N implying the congruence of the Brønsted and Lewis definitions of acidity and basicity in an average sense. A larger value of $\Delta$N indicates a greater amount of electron transfer and hence a better Lewis acid-base pair. Since the base remains same for all the acid-base pairs studied here and it is a very strong base, a larger $\Delta$N would imply a stronger acid and in case it corresponds to a smaller p$K_a$ value (stronger Brønsted acid) these two definitions would not contradict. It is heartening to note that on an average this is true. An arbitrary pair of acids may not always follow a larger $\Delta$N – smaller p$K_a$ dictum. However, an acid with a small p$K_a$ and a small $\Delta$N, implying that the Brønsted-Lowry and Lewis definitions of acids are at variance with each other, is not common. It is important to mention that this paper is not meant for sophisticated p$K_a$ calculation rather the inherent similarity/dissimilarity between the two definitions is analyzed here.

The present work highlights the correlation between p$K_a$ and $\Delta$N (–$\Delta$E) which are global quantities. However, the acidic behavior is expected to be essentially governed by the functional group (–COOH, –OH etc) present in it. Accordingly the group philicity ($\omega_g^+$) has been considered[25] to be a descriptor for p$K_a$ prediction. In the present work we follow a global-local approach for the molecules containing functional groups (–COOH,

–OH), e.g. carboxylic acids and alcohols. While the global behavior is governed by ΔN (–ΔE), the local aspect is taken care of by $\omega_g^+$. Two different series (carboxylic acids and alcohols) are considered for this purpose. Table 2 presents the p$K_a$ values estimated with the two parameter linear regression model in terms of ΔN (–ΔE) and $\omega_g^+$ (both MPA and HPA). A high degree of correlation is observed (Figures 5 and 6) between the calculated and estimated p$K_a$ values with coefficient of correlation ($R^2$), variance adjusted to degrees of freedom ($R_{ADJ}^2$) and variance of leave-one-out cross-validation ($R_{CV}^2$) greater than 0.98 in all cases. The carboxylic acids and the alcohols fall on the same line (slope close to unity and intercept close to zero). However, they fall on different regions of the line helping us to identify the different sets of functional groups. The trend is similar for MPA and HPA calculations. A possible correlation of these descriptors with the cation releasing/ anion accepting power of acids will allow us to develop a generalized acid-base theory encompassing redox and electrophile-nucleophile reactions as well.

## Conclusions

The Brønsted-Lowry and Lewis definitions of acids and bases are in general compatible to each other in the sense that a strong Brønsted acid is generally a strong Lewis acid as well. However, for an arbitrary pair of acids, a stronger Brønsted acid need not necessarily be a stronger Lewis acid. The fractional number of electron transfer between an acid and a base and the energy lowering associated with that process may be considered to be reasonable indicators of the corresponding p$K_a$ values. The situation improves when the local information in terms of the group philicity is also injected into this regression analysis.

## Acknowledgements

We are thankful to BRNS, Mumbai for financial assistance. One of us (K.G.) is thankful to the Indian Academy of Sciences, Bangalore for the Summer Research Fellowship and the Principal, Ramananda College, Bishnupur, Bankura for the grant of study leave to pursue this work at IIT Kharagpur.

**Table 1.** Electronegativity (χ), chemical hardness (η), ΔN and (–ΔE) values with experimental and predicted p$K_a$ values in gas phase

| No | Molecules | χ (eV) | η (eV) | ΔN | (–ΔE) (a.u.) | Exptl p$K_a$[†] | Calcd p$K_a$[†] ΔN | Calcd p$K_a$[†] (–ΔE) |
|---|---|---|---|---|---|---|---|---|
| 1 | Boric acid | 3.4838 | 20.2590 | 0.0278 | 0.0204 | 9.27 | 18.018 | 14.231 |
| 2 | Carbonic acid | 3.9524 | 7.4485 | 0.0710 | 0.0688 | 6.35 | 5.4053 | 4.4793 |
| 3 | Chlorous acid | 4.6785 | 5.2243 | 0.1167 | 0.1554 | 1.94 | 0.9115 | 1.3865 |
| 4 | HClO3 | 5.5948 | 6.0571 | 0.1462 | 0.2616 | -1.00 | -0.2088 | 1.1006 |
| 5 | Perchloric acid | 5.8344 | 6.3238 | 0.1526 | 0.2914 | -1.6 | -0.3635 | 1.0918 |
| 6 | Hydrofluoric acid | 5.3726 | 10.323 | 0.1017 | 0.1707 | 3.2 | 1.8801 | 1.2815 |
| 7 | Nitrous acid | 4.9886 | 6.1573 | 0.1204 | 0.1790 | 3.25 | 0.7206 | 1.2406 |
| 8 | Nitric acid | 5.5313 | 6.7245 | 0.1361 | 0.2393 | -1.3 | 0.0852 | 1.1139 |
| 9 | Sulfamic acid | 4.2950 | 6.2341 | 0.0918 | 0.1046 | 1.05 | 2.7458 | 2.3321 |
| 10 | Sulfuric acid | 4.7737 | 7.1697 | 0.1032 | 0.1424 | 1.99 | 1.7604 | 1.5181 |
| 11 | Sulfurous acid | 4.036 | 6.8140 | 0.0777 | 0.0785 | 1.85 | 4.3971 | 3.6716 |
| 12 | Thiosulfuric acid | 4.4017 | 5.7574 | 0.0999 | 0.1192 | 0.6 | 2.0208 | 1.9142 |
| 13 | Cyanic acid | 4.5103 | 7.2665 | 0.0927 | 0.1157 | 3.7 | 2.6526 | 2.0000 |
| 14 | Thiocyanic acid | 4.5453 | 6.1180 | 0.1028 | 0.1300 | -1.8 | 1.7945 | 1.6972 |
| 15 | Acetaldehyde | 3.8249 | 6.1850 | 0.0731 | 0.0662 | 13.57 | 5.0705 | 4.7384 |
| 16 | Water | 3.9391 | 8.4246 | 0.0658 | 0.0633 | 13.995 | 6.3131 | 5.0410 |
| 17 | H3PO2 | 3.8783 | 6.9477 | 0.0709 | 0.0660 | 2.00 | 5.4243 | 4.7498 |
| 18 | Phoshorous acid | 3.9854 | 7.1199 | 0.0740 | 0.0729 | 1.3 | 4.9333 | 4.1102 |
| 19 | Phosphoric acid | 4.2537 | 6.8625 | 0.0858 | 0.0960 | 2.16 | 3.3821 | 2.6718 |
| 20 | Cyanamide | 3.6121 | 6.9129 | 0.0609 | 0.0487 | 1.1 | 7.2783 | 7.0470 |
| 21 | Acetamide | 2.9856 | 6.5634 | 0.0380 | 0.0185 | 15.1 | 13.718 | 14.959 |
| 22 | Hydrogen peroxide | 3.6378 | 7.5750 | 0.0589 | 0.0478 | 11.62 | 7.7074 | 7.1881 |
| 23 | Hydrogen sulfide | 3.5688 | 6.8096 | 0.0598 | 0.0464 | 7.05 | 7.5296 | 7.4320 |
| 24 | Hydrazoic acid | 4.2806 | 6.4154 | 0.0899 | 0.1018 | 4.6 | 2.9362 | 2.4337 |
| 25 | Formic acid | 4.0520 | 7.2123 | 0.0760 | 0.0774 | 3.75 | 4.6401 | 3.7541 |
| 26 | Acetic acid | 3.7320 | 6.8078 | 0.0660 | 0.0567 | 4.756 | 6.2723 | 5.8469 |
| 27 | Chloroacetic acid | 4.4120 | 5.9961 | 0.0984 | 0.1179 | 2.87 | 2.1468 | 1.9455 |
| 28 | Fluoroacetic acid | 4.0379 | 6.7473 | 0.0782 | 0.0791 | 2.59 | 4.3313 | 3.6315 |
| 29 | Trichloroacetic acid | 5.0260 | 5.6207 | 0.1275 | 0.1919 | 0.66 | 0.4051 | 1.1936 |
| 30 | Trifluoroacetic acid | 4.8195 | 6.6876 | 0.1089 | 0.1527 | 0.52 | 1.3692 | 1.4100 |
| 31 | Dichloroacetic acid | 4.7790 | 5.7868 | 0.1154 | 0.1595 | 1.35 | 0.9811 | 1.3539 |
| 32 | Thioaceticacid | 4.0478 | 4.9780 | 0.0910 | 0.0925 | 3.33 | 2.8200 | 2.8344 |
| 33 | Propanoic acid | 3.7010 | 6.4244 | 0.0668 | 0.0563 | 4.87 | 6.1298 | 5.8962 |
| 34 | 2-Chloropropanoic acid | 4.1402 | 6.1753 | 0.0859 | 0.0913 | 2.83 | 3.3618 | 2.8935 |
| 35 | 3-Chloropropanoic acid | 4.0231 | 6.3284 | 0.0802 | 0.0805 | 3.98 | 4.0599 | 3.5300 |
| 36 | 2-Methylpropanoic acid | 3.6504 | 6.3573 | 0.0652 | 0.0533 | 4.84 | 6.4367 | 6.3238 |
| 37 | 2-Propynoic acid | 4.6556 | 5.8440 | 0.1097 | 0.1448 | 1.84 | 1.3167 | 1.4896 |
| 38 | 2-Chlorobutanoic acid | 4.0922 | 6.1060 | 0.0845 | 0.0877 | 2.86 | 3.5317 | 3.0846 |
| 39 | 3-Chlorobutanoic acid | 3.9595 | 6.1461 | 0.0788 | 0.0766 | 4.05 | 4.2452 | 3.8123 |
| 40 | 4-Chlorobutanoic acid | 3.8855 | 6.1192 | 0.0760 | 0.0710 | 4.52 | 4.6420 | 4.2718 |
| 41 | 4-Cyanobutanoic acid | 4.0740 | 6.2555 | 0.0827 | 0.0851 | 2.42 | 3.7417 | 3.2334 |
| 42 | Butanoic acid | 3.6007 | 6.5573 | 0.0622 | 0.0493 | 4.83 | 7.0192 | 6.9397 |
| 43 | 4-Hyroxybutanoic acid | 3.4035 | 6.2223 | 0.0559 | 0.0388 | 4.72 | 8.3959 | 8.9342 |
| 44 | Acrylic acid | 4.6286 | 5.7286 | 0.1096 | 0.1433 | 4.25 | 1.3210 | 1.5078 |
| 45 | Pyruvic acid | 4.7897 | 5.1200 | 0.1227 | 0.1702 | 2.39 | 0.6138 | 1.2843 |
| 46 | Oxalic acid | 5.2325 | 5.6279 | 0.1361 | 0.2190 | 1.25 | 0.0854 | 1.1360 |
| 47 | Succinic acid | 3.8908 | 5.7772 | 0.0784 | 0.0735 | 4.21 | 4.3052 | 4.0610 |
| 48 | Malic acid | 3.9938 | 5.869 | 0.0820 | 0.0812 | 3.4 | 3.8250 | 3.4865 |
| 49 | Lactic acid | 3.8859 | 6.3363 | 0.0747 | 0.0698 | 3.86 | 4.8335 | 4.3805 |
| 50 | Maleic acid | 4.7327 | 5.1934 | 0.1194 | 0.1622 | 1.92 | 0.7720 | 1.3341 |

| 51 | Methanol | 3.0943 | 7.4844 | 0.0395 | 0.0213 | 15.5 | 13.207 | 13.903 |
| 52 | Ethanol | 2.9938 | 7.2823 | 0.0363 | 0.0178 | 15.5 | 14.366 | 15.229 |
| 53 | Iso-propanol | 3.0181 | 6.9861 | 0.0381 | 0.0191 | 16.5 | 13.714 | 14.719 |
| 54 | Tertiarybutanol | 3.0490 | 6.7281 | 0.0400 | 0.0207 | 19.2 | 13.011 | 14.122 |
| 55 | Propanol | 2.9577 | 7.2117 | 0.0352 | 0.0166 | 16.2 | 14.815 | 15.711 |
| 56 | 2-butanol | 3.0074 | 6.8835 | 0.0379 | 0.0188 | 17.6 | 13.754 | 14.818 |
| 57 | Methanethiol | 3.1577 | 6.2049 | 0.0461 | 0.0263 | 10.33 | 11.034 | 12.218 |
| 58 | Phenol | 2.9956 | 5.2061 | 0.0430 | 0.0211 | 9.99 | 11.997 | 13.976 |

[†]Experimental data as in ref. 20-24

***Table 2.*** Group philicity index ($\omega_g^+$) of substituted carboxylic acids and alcohols with experimental and predicted p$K_a$ values in MPA and HPA schemes in gas phase

| No. | Molecule | $\omega_g^+$ | | Exptl. p$K_a$[†] | Calcd. p$K_a$ ($\omega_g^+$, $\Delta N$) | | Calcd. p$K_a$ ($\omega_g^+$, $-\Delta E$) | |
|---|---|---|---|---|---|---|---|---|
| | | MPA | HPA | | MPA | HPA | MPA | HPA |
| Carboxylic acids | | | | | | | | |
| 1 | Formic acid | 0.9356 | 0.9925 | 3.75 | 3.3013 | 3.1528 | 3.2678 | 3.1330 |
| 2 | Acetic acid | 0.7385 | 0.7610 | 4.756 | 4.3442 | 4.2734 | 4.2624 | 4.1947 |
| 3 | Chloroacetic acid | 0.8879 | 0.9041 | 2.87 | 2.7989 | 2.7510 | 2.8688 | 2.8257 |
| 4 | Fluoroacetic acid | 0.8675 | 0.8675 | 2.59 | 3.4902 | 3.5181 | 3.4872 | 3.5169 |
| 5 | Trichloroacetic acid | 1.0651 | 1.0674 | 0.66 | 1.2461 | 1.2595 | 1.1950 | 1.2014 |
| 6 | Trifluoroacetic acid | 1.2712 | 1.2504 | 0.52 | 1.0389 | 1.2125 | 1.0108 | 1.1808 |
| 7 | Dichloroacetic acid | 1.0281 | 1.0399 | 1.35 | 1.7533 | 1.7381 | 1.7833 | 1.7689 |
| 8 | Propanoic acid | 0.7260 | 0.7270 | 4.87 | 4.3677 | 4.3670 | 4.3125 | 4.3113 |
| 9 | 2-Chloropropanoic acid | 0.8244 | 0.8230 | 2.83 | 3.4156 | 3.4269 | 3.4690 | 3.4821 |
| 10 | 3-Chloropropanoic acid | 0.8785 | 0.8760 | 3.98 | 3.3871 | 3.4244 | 3.4274 | 3.4678 |
| 11 | 2-Methylpropanoic acid | 0.7074 | 0.6990 | 4.84 | 4.4878 | 4.5169 | 4.4214 | 4.4476 |
| 12 | 2-Propynoic acid | 1.0032 | 1.0162 | 1.84 | 2.0197 | 2.0007 | 2.0777 | 2.0618 |
| 13 | 2-Chlorobutanoic acid | 0.7803 | 0.7761 | 2.86 | 3.6262 | 3.6366 | 3.6771 | 3.6883 |
| 14 | 3-Chlorobutanoic acid | 0.8214 | 0.8214 | 4.05 | 3.6442 | 3.6586 | 3.6864 | 3.7039 |
| 15 | 4-Chlorobutanoic acid | 0.8228 | 0.8191 | 4.52 | 3.7252 | 3.7564 | 3.7596 | 3.7933 |
| 16 | 4-Cyanobutanoic acid | 0.8848 | 0.8809 | 2.42 | 3.2869 | 3.3278 | 3.3399 | 3.3837 |
| 17 | Butanoic acid | 0.6802 | 0.6762 | 4.83 | 4.6807 | 4.6905 | 4.5747 | 4.5808 |
| 18 | 4-Hyroxybutanoic acid | 0.6479 | 0.6441 | 4.72 | 4.9929 | 5.0008 | 4.8373 | 4.8401 |
| 19 | Acrylic acid | 0.7947 | 0.8097 | 4.25 | 2.8043 | 2.7211 | 2.8440 | 2.7600 |
| 20 | Pyruvic acid | 0.9477 | 0.9544 | 2.39 | 1.8332 | 1.8047 | 1.9197 | 1.8906 |
| 21 | Oxalic acid | 1.2138 | 1.2162 | 1.25 | 0.4241 | 0.4671 | 0.2831 | 0.3155 |
| 22 | Succinic acid | 0.5280 | 0.5175 | 4.21 | 4.7586 | 4.7292 | 4.7775 | 4.7431 |
| 23 | Malic acid | 0.7600 | 0.7543 | 3.4 | 3.7758 | 3.7894 | 3.8414 | 3.8562 |
| 24 | Lactic acid | 0.8186 | 0.8079 | 3.86 | 3.7805 | 3.8366 | 3.7914 | 3.8476 |
| 25 | Maleic acid | 0.7828 | 0.7914 | 1.92 | 2.5526 | 2.4760 | 2.6205 | 2.5409 |
| Alcohols | | | | | | | | |
| 26 | Methanol | 0.4234 | 0.4036 | 15.5 | 15.4655 | 15.4639 | 15.4353 | 15.4374 |
| 27 | Ethanol | 0.4018 | 0.3692 | 15.5 | 15.9747 | 15.9463 | 15.9858 | 15.9598 |
| 28 | Iso-propanol | 0.3481 | 0.3031 | 16.5 | 17.4575 | 17.4566 | 17.4619 | 17.4633 |
| 29 | Tertiarybutanol | 0.3054 | 0.2536 | 19.2 | 18.6511 | 18.6390 | 18.6429 | 18.6298 |
| 30 | Propanol | 0.4173 | 0.3845 | 16.2 | 15.5332 | 15.5399 | 15.5523 | 15.5561 |
| 31 | 2-butanol | 0.3495 | 0.3029 | 17.6 | 17.4181 | 17.4543 | 17.4217 | 17.4535 |

[†]Experimental data as in ref. 20-24

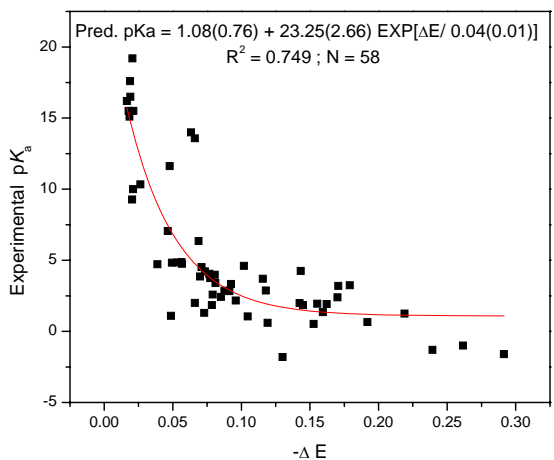
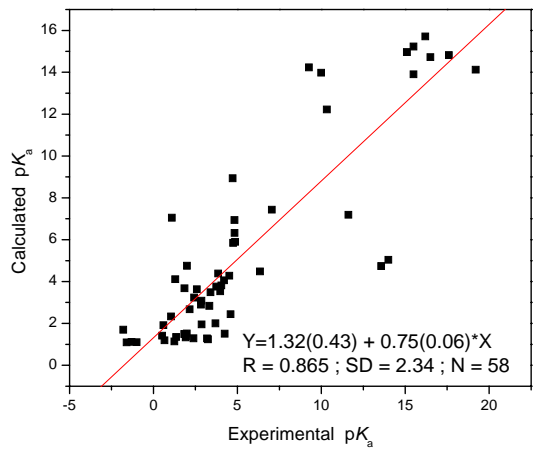

**(a)** **(b)**

*Figure 1.* Relationship between (a) experimental p$K_a$ values of the series of inorganic and organic acids with (–ΔE) in gas phase and their (b) experimental vs calculated p$K_a$ values.

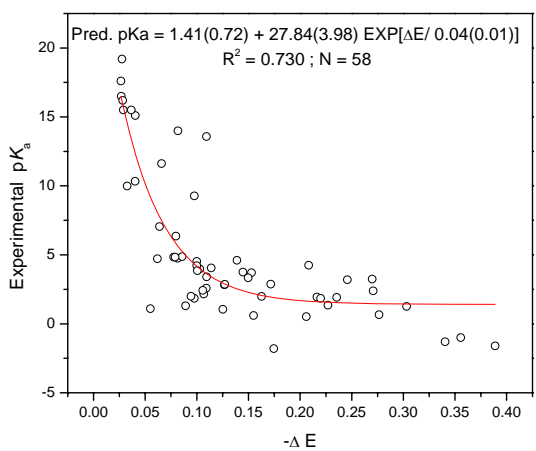
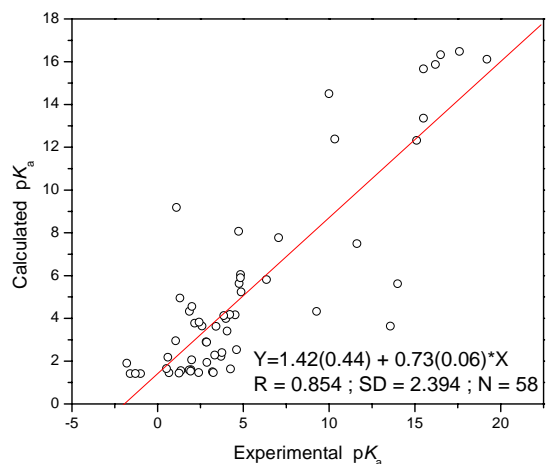

**(a)** **(b)**

*Figure 2.* Relationship between (a) experimental p$K_a$ values of the series of inorganic and organic acids with (–ΔE) in solution phase and their (b) experimental vs calculated p$K_a$ values.

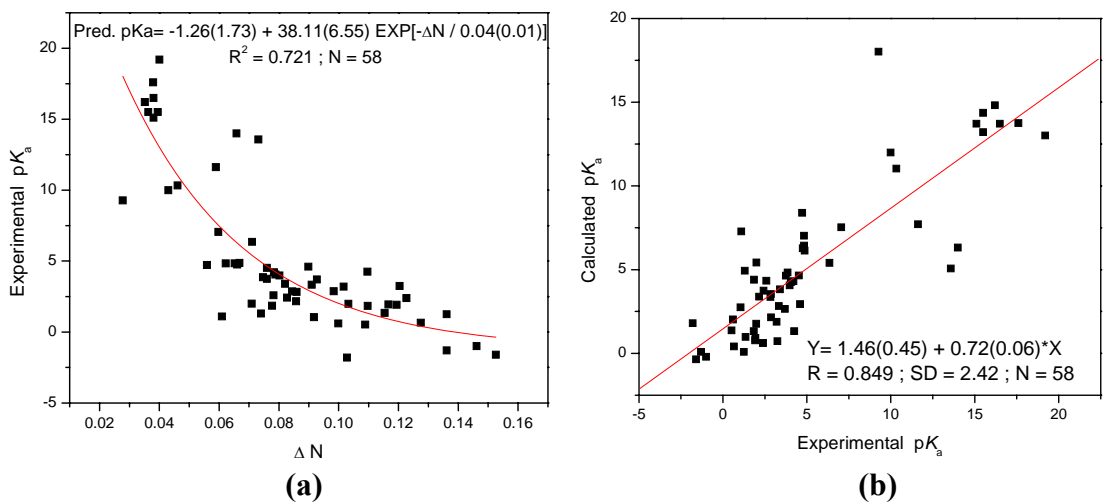

*Figure 3.* Relationship between (a) experimental p$K_a$ values of the series of inorganic and organic acids with ΔN in gas phase and their (b) experimental vs calculated p$K_a$ values.

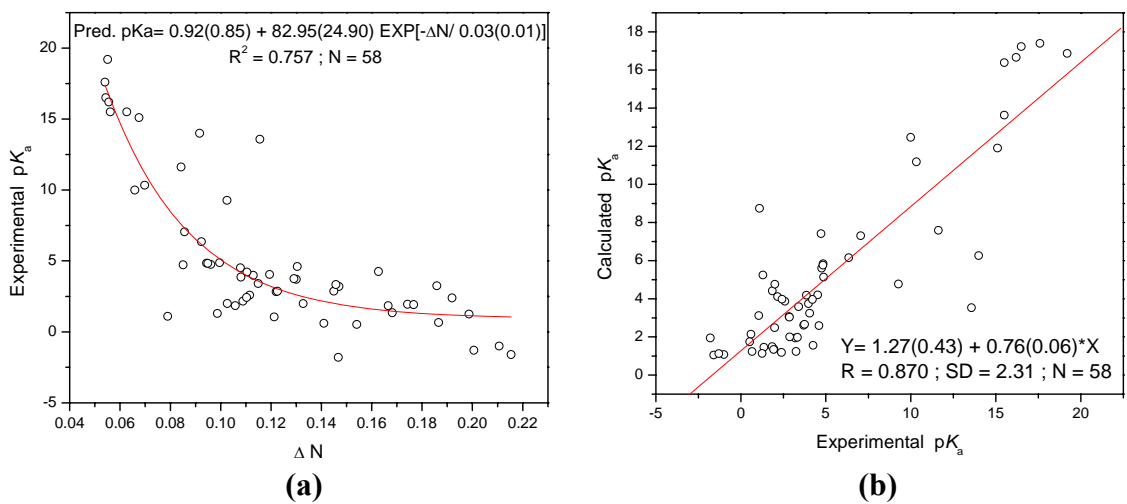

*Figure 4.* Relationship between (a) experimental p$K_a$ values of the series of inorganic and organic acids with ΔN in solution phase and their (b) experimental vs calculated p$K_a$ values.

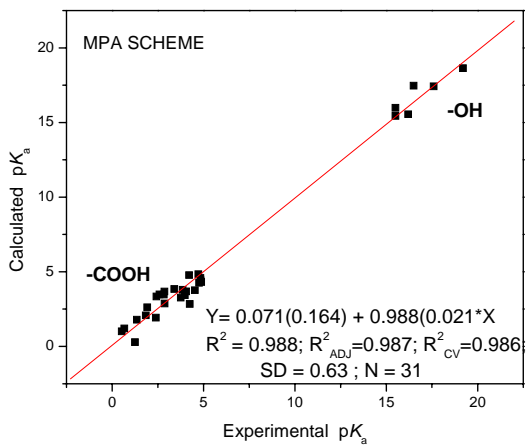 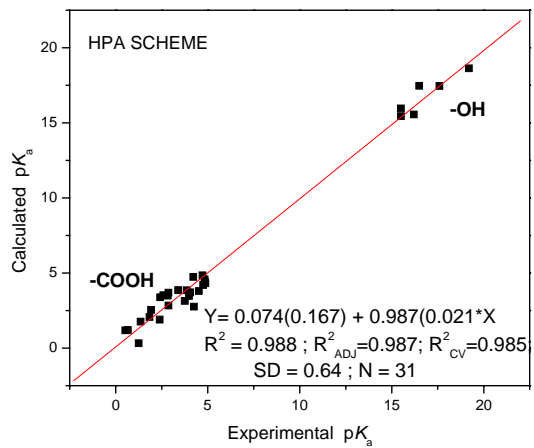

**(a)** **(b)**

*Figure 5.* Relationship between the experimental and predicted p$K_a$ values with $\omega_g^+$ and (-ΔE) of the carboxylic acids and alcohols in a) MPA and b) HPA schemes.

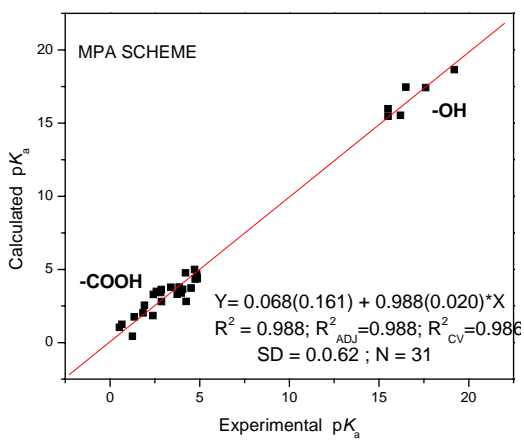 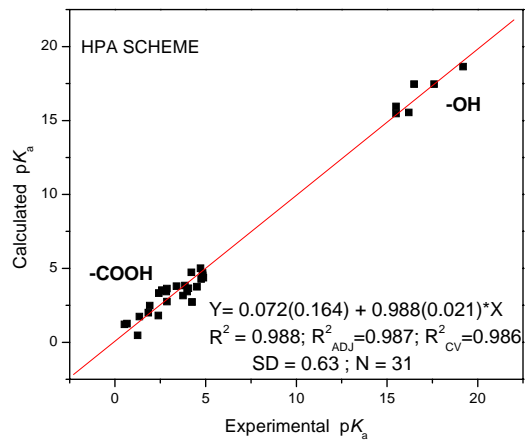

**(a)** **(b)**

*Figure 6.* Relationship between the experimental and predicted p$K_a$ values with $\omega_g^+$ and ΔN of the carboxylic acids and alcohols in a) MPA and b) HPA schemes.